\documentstyle[sprocl]{article}
\begin{document}

\begin{center}

\vspace{5mm}

{\large \bf DENSITY AND CHARGE FLUCTUATIONS IN MULTIPLE PRODUCTION}

\vspace{3mm}

{\large  I.M.\,DREMIN}

\vspace{2mm}

{\normalsize  Lebedev Physical Institute, Moscow 117924, Russia }

\end{center}

\begin{abstract}
{\it Here, I summarize briefly main results of the studies of density and charge
fluctuations in multiparticle production in which I was mainly involved 
during last 3 years.}

The solution of QCD equations for generating functions of {\it parton}
multiplicity distributions reveals new features of cumulant moments
oscillating as functions of their rank. Experimental
data on {\it hadron} multiplicity distributions in $e^{+}e^{-}, hh, hA, AA$
collisions possess the similar features. Contrary, the "more regular" models
like $\lambda \phi ^{3}_6$ predict a different pattern. Evolution of the moments
at smaller phase space bins, zeros of the truncated generating functions
and the singularity of the generating function are briefly discussed.
Such studies can provide some guide to a new representation of multiparticle
processes as compared to the common Fock representation.

Also, the large charge fluctuations can be observed if in some events 
pions tend to be produced in coherent or squeezed isospin states.
\end{abstract}

\section{Introduction and main results}

During the last decade we worked actively on correlations and fluctuations 
of the number of particles produced within small phase space bins inspired
by the ideas of intermittency and fractality. The factorial and cumulant 
moments were especially helpful in this respect. Here I would like to show
that the correlations and fluctuations in large reveal new unexpected features 
in the behavior of the moments as functions of their rank. The larger number
of particles in the total phase space allows to analyze the moments of higher
ranks compared to small bins. This is somewhat different view of the density 
fluctuations within the available phase space volume.

Another interesting problem is related to the charge fluctuations in a single
event. It is closely connected to the so-called Centauro events observed in
cosmic rays.

I present from the very beginning both short review of the problem and the main
results obtained during the last 3 years, leaving their derivation
and brief discussion for the next section.
Those interested in more detailed description (and, in particular, in
formulae  and Figures demonstrating the statements and shown at oral
presentation) should use the list of
references for further reading (e.g., the review papers \cite{1,1a,1b}). The density fluctuations are given by the width (and by higher
moments) of the multiplicity distribution, while charge fluctuations show e.g.
how the share of neutral pions in single events declines from its standard
value of 1/3. The first problem is treated in the framework of QCD, the
second one is considered for coherent and squeezed isospin states.

I would like to stress that QCD predicts the distributions of
partons (quarks and gluons) while in experiment one gets the distributions
of final hadrons. Therefore no {\underline \it quantitative} comparison has
been attempted. To do that, one must rely on the Monte-Carlo models with some 
definite hadronization schemes.
However, the {\underline \it qualitative} features of both
distributions are so spectacular and remind each other that one is tempted
to confirm once again that QCD is a powerful tool for predicting new features
of hadron distributions as well.

For a long time, the phenomenological approach dominated in description of
multiplicity distributions in multiparticle production \cite{2}. The very first
attempts to apply QCD formalism to the problem failed because in the
simplest double-logarithmic approximation it predicts an extremely wide shape
of the distribution in the global phase space \cite{3} (i.e. huge fluctuations)
that contradicts to experimental data. Only
recently it became possible \cite{4} to get exact solutions
of QCD equations for the generating functions of multiplicity distributions
which revealed much narrower shapes and such a novel feature
of cumulant moments as their oscillations at higher ranks \cite{5,6}.
The similar oscillations have been found in experiment for the moments of
hadron distributions \cite{7}. Their pattern differs drastically from those
of the popular phenomenological distributions \cite{1} and of the "non-singular"
(however, possessing the asymptotic freedom property)
$\lambda \phi ^{3}_6$-model \cite{8}. The QCD inspired Monte-Carlo models
describe experiment quite well \cite{8a}.

These findings have several important implications \cite{1}. They show that:

1.{\it the QCD distribution belongs to the class of non-infinitely-divisible ones.}

Two corollaries of this statement follow immediately: \\
{\it a}. The Poissonian cluster models (e.g., the multiperipheral cluster model)
are ruled out by QCD. \\
{\it b}. The negative binomial distribution (so popular nowadays in
phenomenological fits) is not valid. \\
2.{\it the new expansion parameter appears in description of multiparticle
processes.}

Since this parameter becomes large when large number of particles are
involved, it asks for the search of some collective effects and of
more convenient basis than the common particle number (Fock) representation.

QCD is also successful in qualitative description of evolution of multiplicity
distributions with decreasing phase space bins which gives rise to notions
of intermittency and fractality \cite{9,10,11}. The fluctuations
increase in smaller bins. However, there are some new problems
with locations of the minimum of cumulants at small bins \cite{12,13}.

The experimentally defined truncated generating functions possess an intriguing
pattern of zeros in the complex plane of an auxiliary variable \cite{13,14,15}. It recalls
the pattern of Lee-Yang zeros of the grand canonical partition function in
the complex fugacity plane related to phase transition \cite{16,17}
and asks for some collective effects to be searched for \cite{18,1a}.
At high multiplicities these zeros tend to pinch the positive real axis at
the singularity position of the generating function and their study can reveal
the nature of the singularity (see \cite{18,19}) that has
far-reaching consequences for the theory of multiple production.

Besides density fluctuations, there could appear the asymmetry
between neutral and charged pions distributions. The ideas of chiral models,
disoriented chiral condensate, coherent and squeezed isospin states could be
useful in approaching the problem of charge fluctuations. I describe them
briefly in a separate section.

\section{Some QCD technicalities and results}

Let us define the multiplicity distribution
\begin{equation}
P_n = \sigma _n /\sum_{n=0}^{\infty }\sigma _n  ,     \label{1}
\end{equation}
where $\sigma _n$ is the cross section of $n$-particle production processes,
and the generating function
\begin{equation}
G(z) = \sum _{n=0}^{\infty }P_{n}(1+z)^n .     \label{2}
\end{equation}
The (normalized) factorial and cumulant moments of the $P_n$ distribution are
\begin{equation}
F_{q} =\frac {\sum_{n} P_{n} n(n-1)...(n-q+1)}{(\sum_{n} P_{n} n)^q} = \frac {1}
{\langle n\rangle ^q} \frac {d^qG(z)}{dz^q}\vert _{z=0} ,  \label{3}
\end{equation}
\begin{equation}
K_{q} = \frac {1}{\langle n\rangle ^q}\frac {d^q \ln G(z)}{dz^q}\vert _{z=0},
\label{4}
\end{equation}
where $\langle n\rangle = \sum _{n} P_{n} n$ is the average multiplicity. They
describe full and genuine $q$-particle correlations, correspondingly.

Here, I consider QCD without quarks i.e. gluodynamics, since quarks do not
change qualitative conclusions described below. The generating
function of the gluon multiplicity distribution in the global phase-space
volume satisfies the equation
\begin{equation}
\frac {\partial G(z,Y)}{\partial Y} = \int _{0}^{1}dxK(x)\gamma _{0}^{2}
[G(z,Y+\ln x)G(z,Y+\ln (1-x)) - G(z,Y)] .    \label{8}
\end{equation}
Here $Y=\ln (p\theta /Q_{0}), p$ is the initial momentum, $\theta $ is the
angular width of the gluon jet considered,  $p\theta \equiv Q$
where $Q$ is the jet virtuality, $Q_{0}=$const,
\begin{equation}
\gamma _{0}^{2} = \frac {6\alpha _{S}(Q)}{\pi } ,  \label{9}
\end{equation}
$\alpha _S$ is the running coupling constant, and the kernel of the equation is
\begin{equation}
K(x) = \frac {1}{x} - (1-x)[2-x(1-x)] .    \label{10}
\end{equation}
The eq.(\ref{8}) can be solved exactly for fixed coupling constant \cite{6}
and in higher order approximations for the running coupling \cite{4}.
In the last case, the solution of this equation in terms of moments looks as
\begin{equation}
H_q =\frac {K_q}{F_q}=\frac {\gamma _{0}^{2}[1-2h_{1}\gamma +h_{2}(q^{2}
\gamma ^{2}+q\gamma ')]}{q^{2}\gamma ^{2}+q\gamma '}, \label{13}
\end{equation}
where the anomalous dimension $\gamma \approx \gamma _{0}+O(\gamma _{0}^{2})$.
The main prediction is the minimum of $H_q=K_q/F_q$ at 
\begin{equation}
q_{min} =\frac {1}{h_{1}\gamma _{0}}+\frac {1}{2} + O(\gamma _{0}) \approx 5
\end{equation} 
and subsequent oscillations of the ratio $H_q$ at higher $q$.
Let us note that it owes to the singular part of the kernel and is absent 
\cite{8} in more regular theories like $\lambda \phi ^{3}_6$.

While the above results are valid for gluon distributions in gluon jets
(and pertain to QCD with quarks taken into account \cite{1}),
the similar qualitative features characterize the multiplicity distributions
of hadrons in high energy reactions initiated by various particles and nuclei.
The numerous demonstration of it can be found in papers \cite{7,8a}.

Another important feature of the theoretical results is the presence of
the product $\gamma _{0}q$ as a new expansion parameter in all the solutions.
 Formally, it vanishes
in asymptotics. However, since $\gamma _{0}\approx 0.48$ even at $Z^0$-peak
this parameter is large at any rank $q$ and determines the main properties
of the moments. One can say that the asymptotics is unreachable, in practice.
Let us remind that this asymptotics is somewhat similar to the negative binomial
distribution with very wide shape compared to experimental ones (it gives rise
to $H_q =q^{-2}$ while NBD with $k=2$ predicts $H_q =2/q(q+1)$). 
Since at higher ranks one deals with high multiplicity events the product
$\gamma _{0}q$ indicates that for such processes the usual Fock representation
is not convenient. Therefore one is tempted to look for a more suitable
representation for multiparticle processes.

The multiplicity distributions can be measured not only in the total phase
space (as has been discussed above for very large phase-space volumes) but
in any part of it. The most interesting
problem here is the law governing the growth of fluctuations and its possible
departure from a purely statistical behavior related to the decrease of the
average multiplicity in small bins.
Such a variation has to be connected with the dynamics of the
interactions. In particular, it has been proposed \cite{19a}
to look for the power-law
behavior of the factorial moments for small rapidity intervals $\delta y$
\begin{equation}
F_q \propto (\delta y)^{-\phi (q)} \;\;\;\;(\phi (q)>0)\;\;\;\;
(\delta y \rightarrow 0), \label{15}
\end{equation}
inspired by the idea of intermittency in turbulence. In the case of statistical
fluctuations with purely Poisson behavior, the intermittency indices $\phi (q)$
are identically equal to zero.

Experimental data on various processes in a wide energy range support this idea,
and QCD provides a good basis for its explanation as a result
of parton showers \cite{9,10,11}.

Let us turn now to the $q$-behavior of moments at small bins. The phenomenon of
the oscillations of cumulants discussed above reveals itself here as well.
According to the theory \cite{4,12}, the first minimum moves to 
higher ranks at higher
energies because more massive jets become available. Another corollary is that
it should shift to smaller values of $q$ for smaller bins at fixed energy
because the effective value of the anomalous dimension increases due to lower
effective masses of subjets (since $q_{min}\propto \gamma _{0}^{-1}$).
While former statement finds some support in experiment, the second one does
not look to be true (probably, due to higher order terms in the relation above).
Also, one should always keep in mind that lower multiplicities in smaller bins 
prevent from getting higher order moments with good enough precision.

There is another fascinating feature of multiplicity distributions -- it happens
that zeros of the truncated (if the sum in \ref{2} runs up to $n=N_{max}$)
generating function form a spectacular pattern in
the complex plane of the variable $z$. Namely, they seem to lie close to a
single circle. At enlarged values of $N_{max}$ they move closer to the real
axis pinching it at some positive value of $z$.

No QCD interpretation of the fact exists because it is hard to exploit the
finite cut-off in analytic calculations. The interest to it stems from the
analogy to the locations of zeros of the grand canonical partition function
as described by Lee and Yang who related them to possible phase transitions in
statistical mechanics. In that case, the variable $z$ plays the role of
fugacity, and pinching of the real axis implies existence of two phases in
the system considered. The Feynman-Wilson liquid analogy can be used when 
applying this idea to particle production. However, in my opinion, it would be 
premature to consider this phenomenon as a signature of any phase transition
in particle collisions.

In particle physics, it shows up the location of the singularity of the
generating function. The number of zeros of truncated generating functions
increases and they tend to move to the singularity point when $N_{max}
\rightarrow \infty $. Since it happens to lie close to the origin, it
drastically influences the behavior of moments (see (\ref{3}), (\ref{4})), and,
therefore, determines the shape of the distribution. It explains also why the
moment analysis is so sensitive to the tiny details of this shape. At the same 
time the stability of the qualitative features of the moments for different
reactions is very impressive and implies some common dynamics. The study of the
singularities is at the very early stage now (new results have been reported
at this Workshop in \cite{19}), and one can only say that the
singularity is positioned closer to the origin in nucleus-nucleus collisions
and it is farthest in $e^{+}e^{-}$ that appeals to our intuitive guess.

Let us discuss possible implications of the results obtained. The very
existence of the new expansion parameter shows that the particle number
representation traditionally used in particle physics becomes inadequate
for multiparticle production. It asks for the search of another approach.
The appealing example of the analogous situation has been provided by
laser physics where the coherent state representation is successful.
However, in multiple production neither coherent nor squeezed states
look quite promising since they do not fit above findings about distributions.
Probably, their weighted averages giving rise to somewhat similar to the
negative binomial distribution would be suitable. Anyway, the singularity of the
generating function can become a starting point for further progress in
that respect. This is an attractive direction for new research.

To conclude, I would like to stress that, once again, QCD demonstrates its
power in predicting new features of {\it particle} distributions when
dealing with {\it parton} distributions.

\section{Notes on charge fluctuations}

Above, we have discussed the charged multiplicity distributions which determine the
density fluctuations of charged pions. We assumed implicitly that the same is
true for neutral pions as well. However, if confirmed, Centauro events in
cosmic rays imply that the strong charge asymmetry can be observed in some
(probably, rather rare) events at very high energies. The perturbative QCD 
is unable to solve the problem since charge asymmetry appears at the 
hadronization stage only. There is no strong charge asymmetry in the common
models. Therefore, one should rely on different dynamics. The analogy to 
photon states can be useful.

It is well known that the soft photons are
produced in a coherent state. If the soft pions tend to be produced in a
coherent (or squeezed due to the non-linear interaction) state and respect
the isospin conservation, then one can show \cite{20,21} (see also
\cite{1b}) that the
strong charge asymmetry should be seen in the individual events with a
noticeable fraction of them having no charged pions at all. This is done
by considering the projections of these states on the states with the definite
value of the isospin. Due to isospin conservation, the total isospin of the 
pion system created at high energy is strongly restricted and should be much 
less than the number of pions. The differential
distribution of the ratio $f$ of the number of neutral pions to the total
number of pions decreases very slowly as $f^{-1/2}$. The similar
situation is typical for the chiral model used for describing the disoriented 
chiral condensate and the "Baked Alaska" scenaria \cite{22,23}. These
approaches favor the creation of squeezed isospin states as well \cite{24}.
However, neither coherent nor squeezed states are able to reproduce the
pattern of moments oscillations described above.
No more room is available here to discuss the problem. For more details
about coherent isospin states, I
refer to the paper \cite{25} presented at this Workshop.

I should apologize that in these notes I described briefly the main statements 
about density and charge fluctuations in particle production leaving aside 
numerous details contained in the papers presented in the list of references.
However, the size of the presentation prevents the detailed review.

\vspace{1mm}

{\large Acknowledgments}

\vspace{1mm}

I am grateful to W. Kittel for inviting me to participate in this workshop
and for financial support.

This work is partly supported by the Russian Fund for Fundamental Research 
(grant 96-02-16347a) and by INTAS grant 93-79.\\
\newpage

{\large References}


\end{document}